\documentclass[prl,superscriptaddress,floatfix,twocolumn,showpacs,amsmath,amssymb]{revtex4}
\usepackage{graphicx}
\usepackage{dcolumn}
\usepackage{bm}
\begin{document}
\title{Soliton binding and low-lying singlets in frustrated odd-legged
$S=\frac{1}{2}$ spin tubes}

\author{Andreas \surname{L\"uscher}}
\affiliation{School of Physics, University of New South Wales,
  Sydney 2052, Australia}
\author{Reinhard M. \surname{Noack}}
\affiliation{Fachbereich Physik, Philipps-Universit\"at Marburg,
  D-35032 Marburg, Germany}
\author{Gr\'egoire \surname{Misguich}}
\affiliation{Service de Physique Th\'eorique, CEA Saclay, 91191
  Gif-sur-Yvette Cedex, France}
\author{Valeri N. \surname{Kotov}}
\author{Fr\'ed\'eric \surname{Mila}}
\affiliation{Institute of Theoretical Physics,
  \'Ecole Polytechnique F\'ed\'erale de Lausanne, BSP, \\ CH-1015
  Lausanne, Switzerland}


\begin{abstract}
Motivated by the intriguing properties of the vanadium spin tube Na$_2$V$_3$O$_7$, we
show that an effective spin-chirality model similar to that of standard Heisenberg odd-legged
$S=\frac{1}{2}$ spin tubes can be derived for frustrated inter-ring couplings, but with
a spin-chirality coupling constant $\alpha$ that can be arbitrarily small. Using density matrix
renormalization group and analytical arguments, we show that, while spontaneous
dimerization is always present, solitons become {\it bound} into low-lying singlets
as $\alpha$ is reduced. Experimental implications for strongly frustrated tubes are  discussed.
\end{abstract}


\maketitle
\begin{figure}
\includegraphics[width=0.4\textwidth]{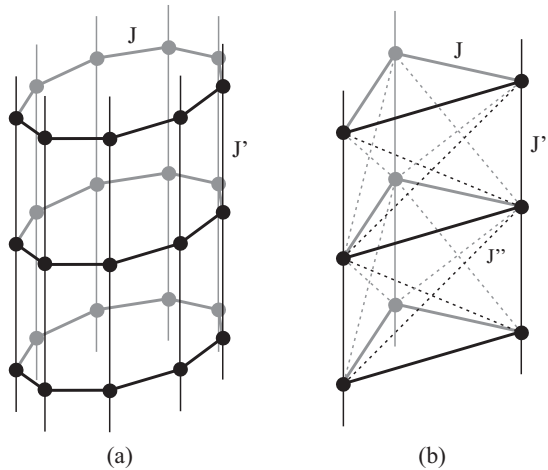}
\caption{\label{fig:models} Two models for Na$_2$V$_3$O$_7$:
(a) non-frustrated, nine-leg spin tube; (b) frustrated, three-leg spin tube.}
\end{figure}

Spin  ladders, systems which consist of a finite number of
coupled chains, have attracted considerable attention recently~\cite{dagotto}.
Spin-$\frac{1}{2}$ ladders with an even number of legs are expected to
have a spin gap, while ladders with an odd number of legs
behave like spin chains at low energy. Both predictions have been
largely confirmed experimentally.
Here we have taken the ladders to have open boundary
conditions in the rung direction, which is the most natural definition.
In comparison, spin ladders with periodic boundary conditions in the
rung direction (see Fig.~\ref{fig:models}), often referred to as spin tubes,
have received much less attention,
mostly because the prospect for experimental realizations was remote.
Nonetheless, it was noticed early on that spin tubes with an odd number of legs
are not expected to behave in the same way as their ladder counterparts.
The crucial observation is that the ground state of a ring
with an odd number of sites is  not just two-fold degenerate,
as for a rung in an odd-leg ladder, but is  four-fold degenerate:
In addition to the Kramers degeneracy of the spin-$\frac{1}{2}$ ground
state, there is a degeneracy due to the two possible signs
of the ground state momentum, leading to an extra degree of freedom on top
of the total spin, often called the chirality by extension of the case of a triangle.
As a consequence, a standard $L$-leg spin tube (Fig.~\ref{fig:models}(a)),
defined by the Hamiltonian
\begin{equation*} H = J \sum_{r,\,l=1}^{N-1,\,L} {{\bm \sigma_{r,\,l}}
    \cdot  {\bm \sigma_{r,\,l+1}}}+ J^\prime \sum_{r,\,l=1}^{N-1,\,L} {{\bm
      \sigma_{r,\,l}} \cdot  {\bm \sigma_{r+1,\,l}}}\;,
\end{equation*}
where ${\bm \sigma_{r,\,l}}$ is a spin-$\frac{1}{2}$ operator on ring $r$
and leg $l$, can be described in the strong-ring limit ($J'\ll J$) by
an effective model, valid to first order in the inter-ring
coupling $J'$~\cite{subrahmanyam, schulz,kawano, cabra},
defined by the Hamiltonian
\begin{equation}
H_\text{\it eff} = K \sum_{r=1}^{N-1} {{\bm S_r}
  \cdot  {\bm S_{r+1}} \left(1 + \alpha \left(\tau_r^+ \tau_{r+1}^- +
  {\rm h.c.} \right) \right)}\;. \label{eq:effectiveham}
\end{equation}
Here ${\bm S_r}$ are the usual spin-$\frac{1}{2}$ operators
which describe the total spin of ring $r$, while ${\bm \tau_r}$ are
pseudo-spin-$\frac{1}{2}$ operators acting on the
chirality. The parameters of the model are an overall coupling
constant $K=\frac{J'}{L}$ and a parameter $\alpha$ that measures the
strength of the coupling between spin and chirality. For ordinary spin
tubes, this coupling is always strong: $\alpha$ is equal to 4 for
three-leg spin tubes and increases with the number
of legs~\cite{kawano,cabra}. Using bosonization arguments~\cite{schulz},
Schulz predicted that the ground state
should be spontaneously dimerized and the spectrum gapped in all sectors, a prediction
supported by further numerical and analytical work~\cite{kawano,cabra,wang}.
In addition, the excitations have been argued to be unbound solitons
and the gap is always a significant fraction of $J'$.
All of this remarkable physics still awaits an experimental realization.

In this context, the recent synthesis of Na$_2$V$_3$O$_7$~\cite{millet}, whose structure
may be regarded as a spin-$\frac{1}{2}$ nine-leg spin tube, has opened up new
perspectives. However, the properties reported so far~\cite{gavilano} do not
match the properties predicted for standard odd-legged spin tubes. In
particular, no spin gap could be detected in zero external field. This
might not be too suprising, however: although
the overall topology of Na$_2$V$_3$O$_7$ is indeed that of a nine-leg spin tube,
the actual geometry is quite different from that of Fig.~\ref{fig:models}(a).
Although ab-initio calculations have not yet reached a consensus~\cite{whangbo, mazurenko},
it is likely that the inter-ring coupling exhibits some kind of frustration.
Since the tubes in Na$_2$V$_3$O$_7$ only have a $C_3$-axis, a frustrated
model of the type of Fig.~\ref{fig:models}(b) might be more appropriate.

In this Letter, we use extensive Density Matrix Renormalization Group (DMRG)
simulations~\cite{white} supported by
several analytical arguments
to show that inter-ring frustration can have dramatic consequences
for the properties of odd-legged spin tubes. In particular, we show that
it can reduce the spin gap and bind solitons in
low-lying singlets. This picture might resolve some of
the puzzles of Na$_2$V$_3$O$_7$.

Our starting point is to notice that, as long as the inter-ring coupling does not
break the rotational symmetry of the tube, the effective Hamiltonian
is still given by Eq.~(\ref{eq:effectiveham}), but with a parameter $\alpha$
that can take on arbitrarily small values if frustration is allowed.
For instance, for the three-leg spin tube of Fig.~\ref{fig:models}(b),
$K=(J'+2 J'')/3$ and $\alpha = 4|J'-J''|/(J'+2J'')$, leading to $\alpha=2$
if each site is coupled to two neighbors (i.e., $J'=0$)~\cite{nojiri}, and to $\alpha=0$
if each site is coupled to all sites of neighboring rings ($J'=J''$).

Therefore, we concentrate on the model of Eq. (\ref{eq:effectiveham})
with $K=1$ and consider all values of $\alpha \geq 0$ in the following.
In a previous DMRG study of the effective model (\ref{eq:effectiveham}),
Kawano and Takahashi~\cite{kawano} reported the finite-size scaling of
the spin gap for the triangular tube  with $\alpha=4$.
Using White's DMRG algorithm~\cite{white} as well, we extend their
numerical analysis to the range $0\le \alpha \le 20$.
For this purpose, we classify the lowest lying excitations according
to the quantum numbers [$S^z$, $\tau^z$] and study the
sectors [0,0], [0,1], [1,0] and [1,1] for  open chains with up to $N=200$ sites.
From an analysis of the truncation-dependence of the gaps, we
find that convergence is reached by keeping 250 states and thus
perform the calculations up to that limit within six finite-system
sweeps.
The sum of the discarded density-matrix eigenvalues is
smaller than $10^{-5}$ in all cases.

\begin{figure}
\includegraphics[width=0.40\textwidth]{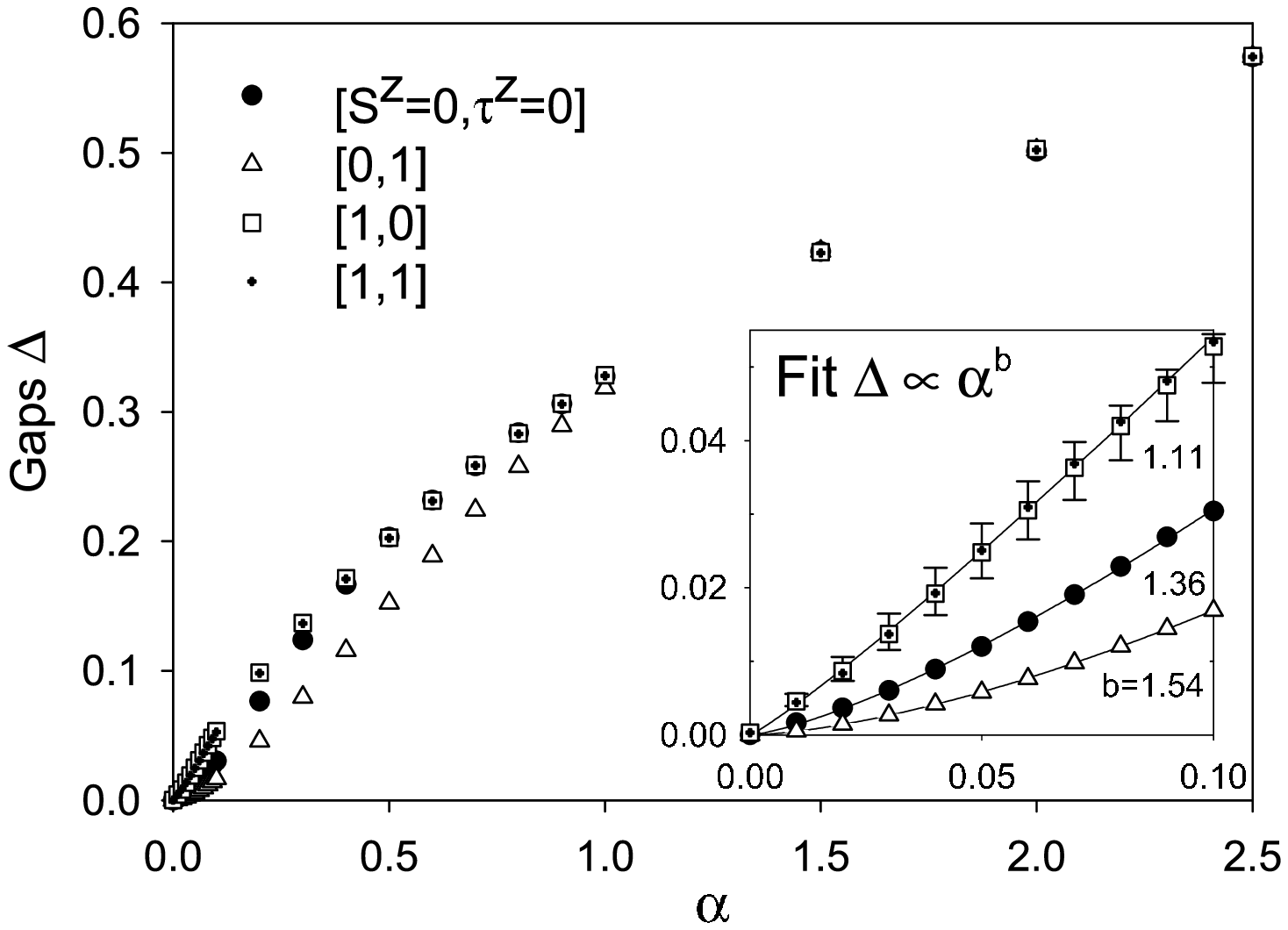}
\caption{\label{fig:gaps}
Extrapolated excitation gaps in the four [$S^z$, $\tau^z$] sectors as a
function of $\alpha$. Inset: Power-law fits of the small $\alpha$ results.}
\end{figure}

Above $\alpha\simeq 1.4$, a range that includes all non-frustrated
spin tubes, we found that a gap is indeed present, in agreement with
Kawano and Takahashi's analysis of the case  $\alpha=4$,
but interestingly enough, we find that the first excitation appears in all sectors.
In other words, spin and chirality gaps are equal above $\alpha\simeq 1.4$.

The situation changes dramatically upon reducing $\alpha$. Below $\alpha\simeq 1.4$,
the first excitation is no longer degenerate but appears only in the sector
[$S^z$=0, $\tau^z$=1]. The first excitation is thus a chirality excitation, and
the spin gap $\Delta_S$ and the chirality gap  $\Delta_\tau$ are no longer equal.
To examine this point further,
we have performed a systematic analysis for $\alpha<1.4$ in all
sectors [0,0], [0,1], [1,0] and [1,1], including careful finite-size
scaling. The first excitation is always in the sector
[$S^z$=0, $\tau^z$=1] and is non-degenerate.
Below $\alpha\simeq 0.5$,
the second excitation is also non-degenerate and appears in the sector
[0,0]. In this parameter range, the first excitation that has a nonzero
spin quantum number is the third excited state.
This excitation manifests itself in all sectors.
By following this excitation as $\alpha$ is increased,
one can determine that it becomes the second excitation
at $\alpha\simeq 0.5$, and then the first excitation at $\alpha \simeq 1.4$.

The gaps corresponding to these excitations are plotted in Fig.~\ref{fig:gaps}.
In extracting the gaps, some care had to be taken regarding finite-size effects.
The results were fitted with polynomials in $1/N$, where $N$ is the
length of the tube. Good fits could be obtained with third order polynomials.
The extrapolated values lie between those obtained with quadratic
and quartic polynomials. The differences between these fits
were used to define the error-bars shown
in the inset of Fig.~\ref{fig:gaps}.

That these gaps correspond to very different excitations is confirmed
by their $\alpha$-dependences.
As shown in the inset, the results for $\alpha \le 0.1$ can be fitted with
power laws of the form $\Delta \propto \alpha^b$ with
exponents $b=1.54\pm0.06$, $1.36\pm0.07$,  and $1.11\pm0.12$
for the singlet excitations in the sectors
[0,1], [0,0], and for the first spin excitation, respectively.
These exponents are consistent with the simple fractions
$b=3/2$, $4/3$ and $1$.
The large error bar and the value significantly larger than $b=1$
for the spin excitation
is very probably a finite-size effect. Since the spin gap follows a
linear finite-size scaling for spin tubes with up to 200 rings
when $\alpha$ is very small, the extrapolations
underestimate the gap, resulting in an exponent larger than the actual one.

\begin{figure}
\includegraphics[width=0.48\textwidth,clip]{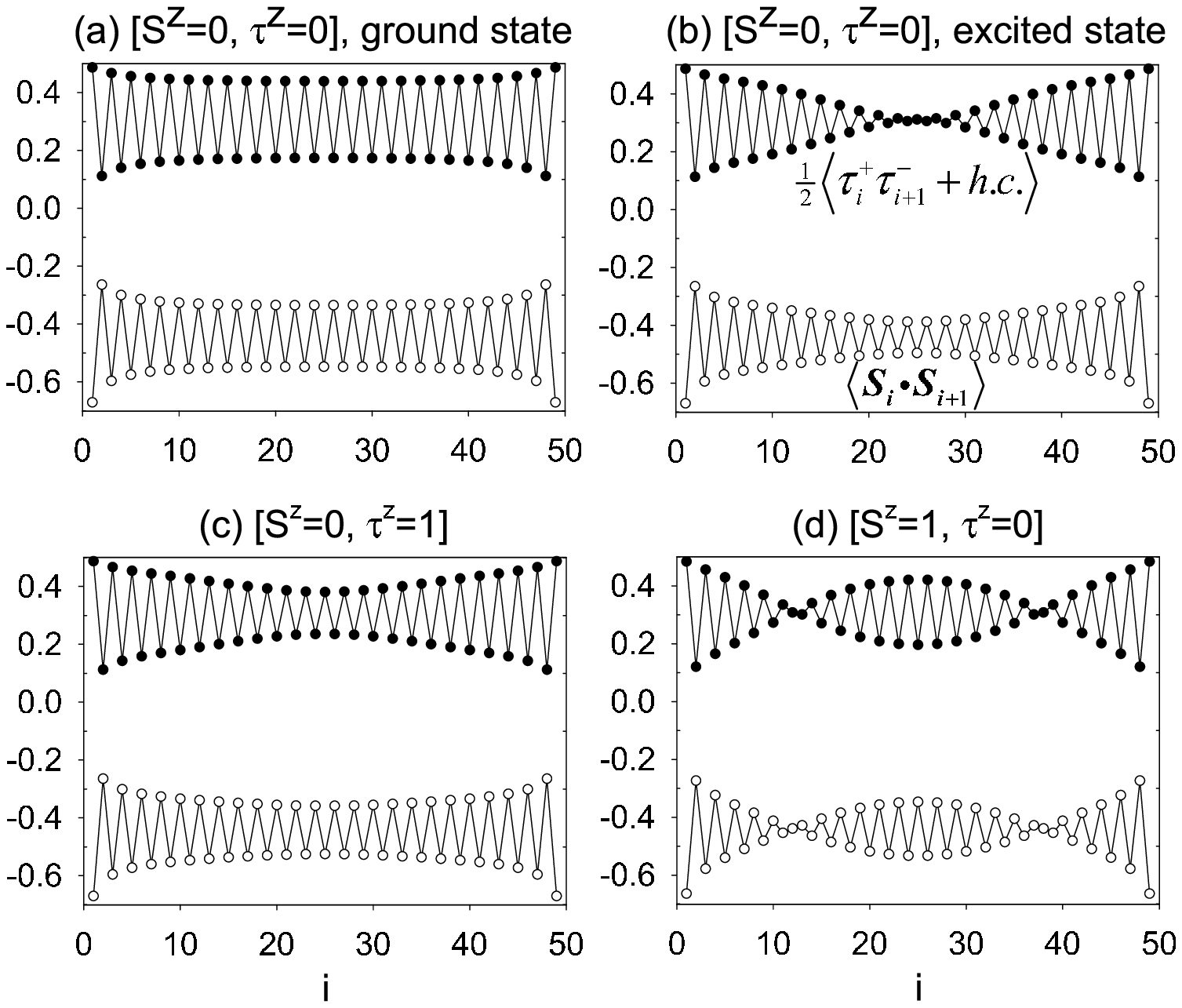}
\caption{\label{fig:correlations}
Nearest-neighbor expectation values of spin and chirality along a
chain with 50 sites in the sectors [$S^z$= 0, $\tau^z$= 0], [0, 1]
and [1, 0] for $\alpha=0.1$.
The ground state (a) is dimerized.
The unaltered dimer patterns and the constrictions at the center of
the chain in (b) and (c) indicate bound soliton states,
whereas the two domain walls in (d) suggest the presence of two
unbound solitons.}
\end{figure}

The nature of the excitations can be further explored by examining
the nearest-neighbor expectation values of the spin and pseudo-spin
interactions
$\left\langle {\bm S_i} \cdot {\bm S_{i+1}} \right\rangle$ and
$\left\langle \tau_i^+ \tau_{i+1}^- + \text{h.c.} \right\rangle$.
In Fig.~\ref{fig:correlations},
these quantities are shown for
the lowest-lying states in the
important [$S^z$, $\tau^z$] sectors.
The spin and chirality degrees of freedom alternate synchronously.
As expected,
the ground state, shown in Fig.~\ref{fig:correlations}(a), is
uniformly dimerized in both the spin and the chirality channels.

Excitations in a dimerized spin chain can be described in terms of
solitons, which can be viewed as domain walls
between two dimer coverings~\cite{majumdar,shastry}.
In a chain with an even number of sites, solitons always appear in
pairs, which can either be bound or unbound.
The bound states can also be interpreted as excited bonds, i.e.,
as spin-triplet states in a background of dimers,
and would therefore leave the
dimer pattern unchanged.
In an open chain, a bound soliton state is located with maximum
probability at the center of the system,
whereas its unbound counterpart tries to maximize
both the distance between the solitons and the distance to the ends
of the chain.

The nearest-neighbor expectation values of excited states in the
sectors [$S^z$= 0, $\tau^z$= 0] and [0, 1] for $\alpha=0.1$ are shown
in Fig.~\ref{fig:correlations}(b) and (c).
The unaltered dimer patterns and the constrictions at the center of
the chain indicate the presence of bound soliton states.
The very pronounced constriction in Fig.~\ref{fig:correlations}(b)
suggests that the excitation in the
ground state sector is close to unbinding, which
we find to occur in the parameter range
$0.5 < \alpha < 0.6$.
For larger values of $\alpha$,
all expectation values have a structure similar to
that shown in Fig.~\ref{fig:correlations}(d), representing the
lowest-lying spin excitation.
Here one can clearly identify two domain walls at around $\tfrac{1}{3}$
and $\tfrac{2}{3}$ of the chain length, suggesting the presence of two
unbound solitons.

We have checked that
these features are independent of the
chain length by studying systems of up to 200 sites.
From a complete analysis of the expectation values throughout the
whole range of $\alpha$, we
conclude that lowest-lying bound
states only exist in the sectors [$S^z$= 0, $\tau^z$= 0] and [0, 1],
for which the unbinding can be observed at $0.5 < \alpha < 0.6$ and
$1.3 < \alpha < 1.4$, respectively.
The lowest-lying pure spin and combined spin-chirality excitations are
always unbound and are therefore degenerate in the thermodynamic limit.

Interestingly, the same hierarchy of states for the
bound and unbound soliton excitations as a function of $\alpha$ can be obtained
analytically  by
the variational approach
of Wang~\cite{wang} for the model that includes a next-nearest-neighbor
coupling along the legs with relative strength $\beta=0.5$~\cite{luescher}.
Since it was shown by Kawano and Takahashi~\cite{kawano} that Wang's
model remains in
the same phase as the additional interaction $\beta$ is turned off
for $\alpha=4$,
we conjecture that the two models are in the
same phase over the whole range of $\alpha$.
In the same spirit, we note that
the effective Hamiltonian of Eq.~(\ref{eq:effectiveham}) can also be seen as a
special case of the recently
investigated spin-orbital
models~\cite{mila,kolezhuk,yamashita,pati}, with great similarities
in the roles of  chirality and orbital degrees of freedom.

Next, we show that
the scaling of the chirality gap $\Delta_\tau \propto \alpha^{3/2}$ can be recovered by
a simple mean-field decoupling of the interaction
terms ${\bm S_r} \cdot  {\bm S_{r+1}} \tau_r^\pm \tau_{r+1}^\mp$
into
\begin{equation}
\langle {\bm S_r} \cdot  {\bm S_{r+1}} \rangle \tau_r^\pm \tau_{r+1}^\mp
+{\bm S_r} \cdot  {\bm S_{r+1}} \langle \tau_r^\pm \tau_{r+1}^\mp\rangle
-\langle {\bm S_r} \cdot  {\bm S_{r+1}}\rangle \langle \tau_r^\pm
\tau_{r+1}^\mp \rangle \;.
\nonumber
\end{equation}
Since, according to our numerical results, the ground state is
spontaneously dimerized, we look for a dimerized solution
by starting with alternating expectation values
$\langle  {\bm S_r} \cdot  {\bm S_{r+1}}\rangle
= C_S-\left(-1\right)^r \delta_S $ and
$\langle  \tau_r^+ \tau_{r+1}^- + \tau_r^- \tau_{r+1}^+\rangle
= C_\tau-\left(-1\right)^r \delta_\tau $,
where
$\delta_{S/\tau}$ is the alternation parameter in the corresponding
channel.
The mean-field Hamiltonian then describes
Heisenberg and XY-chains with alternating bond strengths.
Now, for
Heisenberg and XY-chains with alternating exchange  $J[1+(-1)^r \epsilon]$,
the scalings of the gap and of the alternation parameters
as a function of $\epsilon$
are well known~\cite{cross,black,affleck,pincus}. Up to logarithmic corrections,
they are given by:
$\Delta_S\propto J(\epsilon)^{2/3}$, $\delta_S\propto \epsilon^{1/3}$,
$\Delta_\tau \propto J\epsilon$ and $\delta_\tau \propto \epsilon$.
The mean-field decoupling then leads to a $J$ of order one and
$\epsilon \propto \alpha \delta_\tau$ for the spin part, and to
$J \propto \alpha$ and $\epsilon \propto \delta_S$ for the chirality.
Self-consistency then requires that
$\delta_S,\delta_\tau\propto \alpha^{1/2}$
and $\Delta_\tau \propto \alpha^{3/2}$. This last scaling is in very good
agreement with our DMRG result for the chirality gap, for which no
logarithmic correction could be extracted within our numerical accuracy.
As a further check, we have also extracted $\delta_S$ and $\delta_\tau$
as a function of $\alpha$. They are not strictly proportional, but we
believe this is due to logarithmic corrections.
Indeed, the ground state energy for small $\alpha$ reads
\begin{equation} \label{eq:meanfieldenergy} \begin{split}
e_\text{\it eff}^{MF} = \; & A \alpha \delta_S^2
\left(1+B \log{\delta_S}\right)+ \\ &C \left(\alpha \delta_\tau \right)^{4/3}
\left(\log{D \alpha \delta_\tau}\right)^{-1}
-\alpha \delta_S \delta_\tau + \text{const.}
\end{split} \end{equation}
$A$, $B$, $C$, and $D$ are
constants~\cite{orignac} independent of $\alpha$.
The alternation parameters $\delta$ minimizing the ground state energy
obey the relation
$\delta_\tau \propto \delta_S (\log{\delta_S}+{\rm const.})$,
leading to a very good fit of our numerical results (not shown).

\begin{figure}
\includegraphics[width=0.40\textwidth,clip]{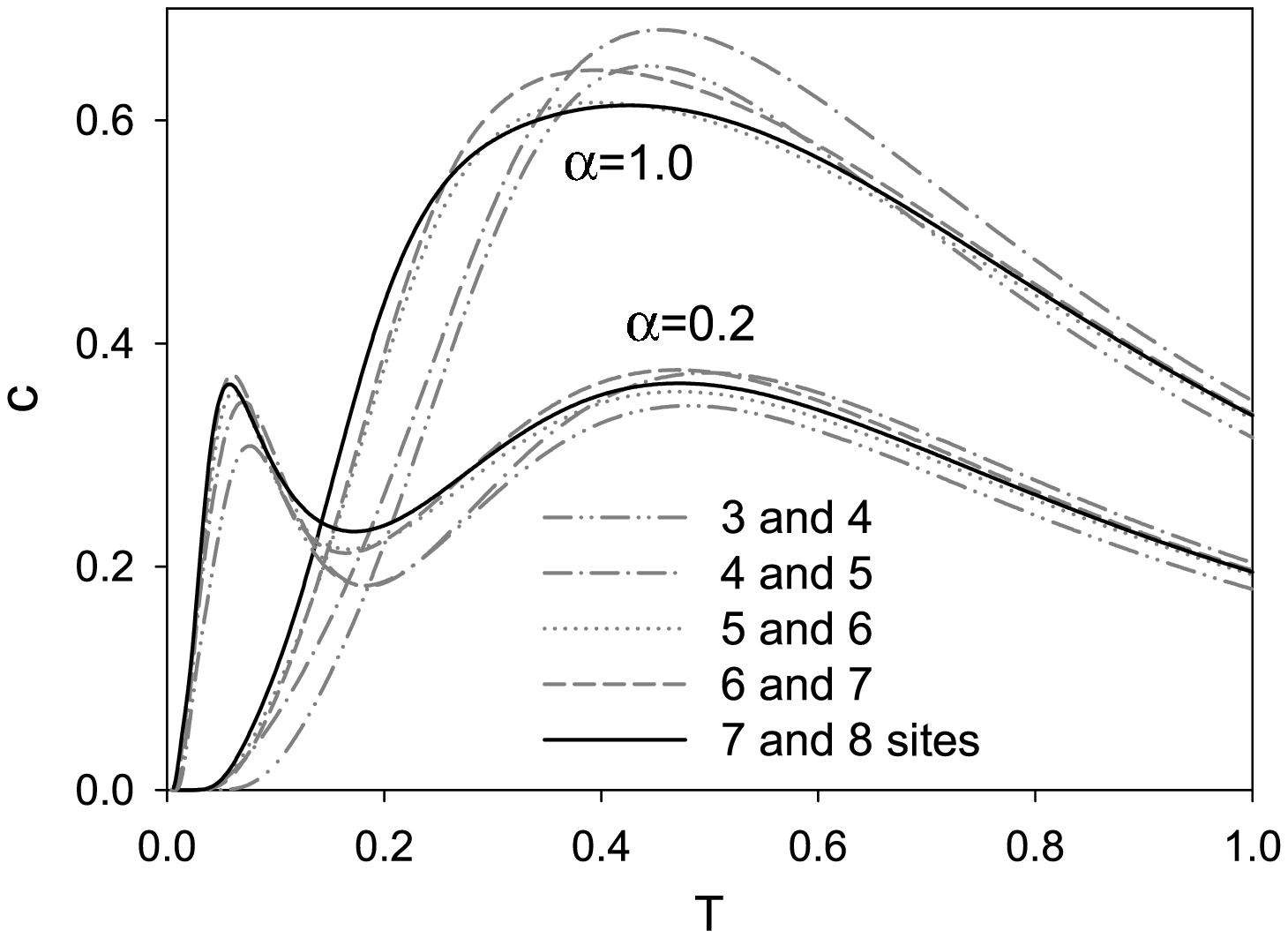}
\caption{\label{fig:specheat}
  Specific heat per site as a function of the temperature $T$ for
  $\alpha=0.2$ and $1.0$ calculated by exact diagonalization of
  systems with up to 8 sites.}
\end{figure}

In order to illustrate the influence of the chirality degrees of freedom,
we have calculated the specific heat density $c$ as a function of
temperature using exact diagonalization of small systems.
We have used weighted means between systems with an even and odd number of
sites~\cite{bonner} to extrapolate the curves to the bulk limit.
Extrapolations for systems with up to 8 sites and two particular
values of $\alpha$ are shown in Fig.~\ref{fig:specheat}.
The double-peak structure for small $\alpha$ is due to the
well-separated low-lying chirality excitations.
The distinct low-temperature peak progressively disappears as the
chirality excitations get closer to the spin excitations and is
completely absent at $\alpha\approx 1$. We also note that these
low-lying chirality excitations, which are singlets, should be detectable
in Raman spectroscopy. Both predictions are expected to apply to
Na$_2$V$_3$O$_7$ if the absence of any detectable zero-field spin gap
in that compound is indeed a consequence of frustration.

In conclusion, we have shown that the low energy properties
of odd-legged spin tubes change dramatically if the coupling
between spin and chirality degrees of freedom is reduced, as
would be the case for frustrated spin tubes. While our results
confirm that excitations in non-frustrated
or weakly frustrated spin tubes are
unbound solitons, they show that solitons are bound
into singlet bound states if the coupling between spin and
chirality is sufficiently reduced by frustration. As a consequence,
spin and chirality gaps are no longer equal, and the low-lying
chirality excitations, which lie below
the spin gap, are expected
to lead to a number of interesting consequences such as
an additional low-temperature peak in the specific heat. We hope
that these conclusions will encourage further experimental
investigations of Na$_2$V$_3$O$_7$ and of other odd-legged spin
tubes~\cite{nojiri}.

\begin{acknowledgments}
It is a pleasure to thank J. Gavilano, H.~R.~Ott, J.~Sirker and
O.~Sushkov  for valuable discussions.
G.~M.\ acknowledges the hospitality
of IRRMA (Lausanne), and V.~K.\ and F.~M.\ acknowledge the support of
the Swiss National Fund.
\end{acknowledgments}

\end{document}